# Neutron and Muon Flux Measurements at BEO Moussala towards to Space Weather Research[*]

Alexander Mishev on behalf of BEO Moussala

*Institute for Nuclear Research and Nuclear Energy, Bulgarian Academy of Sciences, Basic Environmental Observatory Moussala 72 Tsarigradsko chaussee, Sofia 1284, Bulgaria*

Corresponding author: A. Mishev E-mail: mishev@inrne.bas.bg Tel: ++35929746310 Fax: +35929753619

**Abstract:** *The existing and in development devices at Basic Environmental Observatory (BEO) Moussala are presented, precisely the experiments connected with space weather and astroparticle studies. The recent results of the secondary cosmic ray measurements using atmospheric Cherenkov light telescope are presented precisely the possibility to study the atmospheric transparency. The final design of neutron flux meter based on SNM-15 detectors is described as well several preliminary experimental and Monte Carlo results. The developed muon telescope based on water Cherenkov detectors is also presented. The scientific potential is discussed, precisely the connection between cosmic ray measurements and the environmental parameters, precisely atmospheric parameters. The project for muon hodoscope and multidirectional Nor Amberd neutron monitor with the preliminary studies is shown, precisely several estimations and the possible design.*

## 1. Introduction

During the last decades the high mountain observatories have been exploited not only for cosmic ray and astroparticle studies but for environmental studies and observations of the Sun-Earth system. In this connection the Basic Environmental Observatory BEO Moussala (the general view of the station in shown Fig. 1) located on the top of the highest mountain at Balkan Peninsula, precisely at 2925m above sea level is a privileged place for such type of investigations.

This is one of the most proper places in the region of Balkans with relatively small anthropogenic influence and therefore gives excellent possibility for high-mountain monitoring i.e. possibilities for monitoring for changes and processes in the atmosphere, as example the air-transport, aerosol and UV measurements, changes of gamma-background,

---

dose-rate, trans-border pollutants transfer etc… At the same time the BEO Moussala gives possibility to exploit the secondary cosmic ray neutron and muon flux, the atmospheric Cherenkov light in attempt to study different problems of cosmic rays, space weather and connections of the Sun-Earth system.

The detailed analyses of the collected data give information about possible relation between very different kind of parameters and factors. Generally the following specific objectives are pursued in attempt to provide basic information permitting analysis of the connection between cosmic ray variation and several atmospheric parameters. The aim is the detailed, precise and contemporary measurements of cosmic ray intensity especially the muon, electron, gamma and neutron component and the atmospheric Cherenkov light. At the same time the atmosphere parameters, including anthropogenic products as well in different conditions of altitude, latitude and urban development is needed.

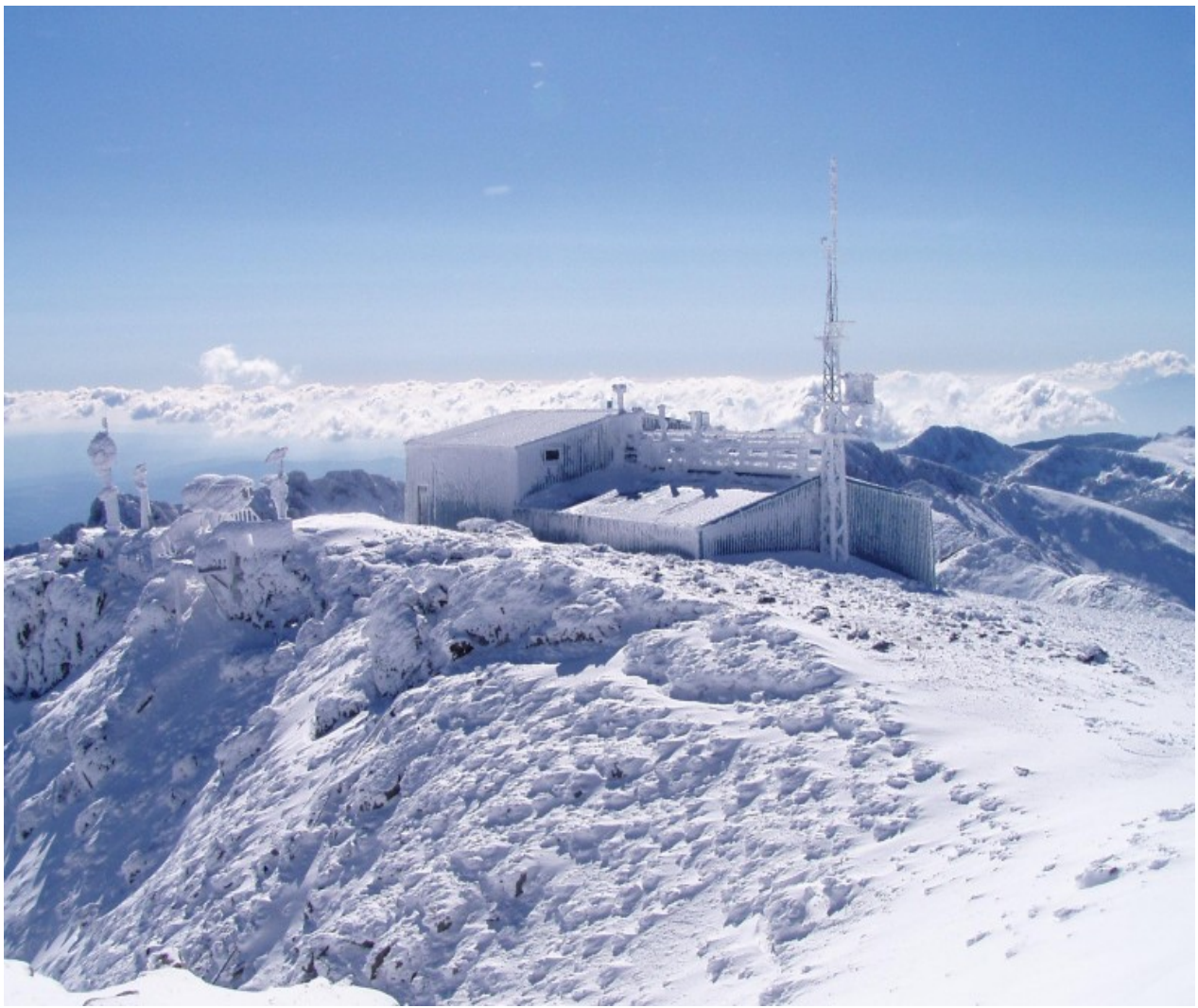

Fig.1 General view of Basic Environmental Observatory Moussala

In the last years one of the most existing topics in the area of Sun-Earth system investigations is the possible influence of cosmic ray on terrestrial atmosphere, precisely the connection between low energy cosmic ray and the Earth atmosphere. As example the variations of the cosmic rays may be responsible for the changes in the large-scale atmospheric circulation associated with solar activity phenomena [1, 2]. Presently several arguments claims that the Solar activity affects the global climate in different aspects and timescales. One possibility is based on climate response to changes in the cosmic ray flux and radiative budget [3]. This is connected with the tropospheric response to solar variability precisely the heating of the troposphere during solar maximum. This is related with the modulation of the large-scale tropospheric circulation systems [4]. Additionally the stratospheric ozone plays important role on the modulation of the radiative influence of the climate. [5]. A powerful tool for investigations from Earth the variation of cosmic ray flux is based on registration of secondary cosmic ray neutrons and muons [1]. Moreover among the different proposed mechanisms as example the UV heating of the stratosphere [2] or change of the solar irradiance [6] the influence of cosmic ray to cloud formation [7, 8] seems to be most promising at experimental point of view and existing and in development devices at BEO Moussala.

Therefore the precise and contemporary measurements of secondary cosmic ray neutrons and muons are very important. At the same time such type of measurements gives good basis for study of solar-terrestrial influences and space weather. Moreover the ability to forecast for long term space weather needs a precise knowledge of solar activity.

The space weather refers to conditions on the sun, solar wind and Earth's magnetosphere and ionosphere [9]. Several characteristic signatures in cosmic ray may be used for space weather applications [10] on the basis of secondary cosmic ray neutron data. Good examples are the solar proton events and Geomagnetic storms.

One of the significant points is related with the atmospheric transparency and cloud formation. As was mentioned above the variations of the cosmic rays, both solar and galactic may be responsible for the changes in the large-scale atmospheric circulation. It is possible to associate such type of phenomena with solar activity and precisely with the energy of cosmic particles being 0.1–1 GeV [7, 11]. It is obvious that possible mechanism of cosmic ray effects on the lower atmosphere involves changes in the atmospheric transparency which is connected with cloud cover. This is possibly due to the changes in the stratospheric ionization produced by the considered cosmic particles. Moreover cosmic ray reflects on the atmospheric temperature assuming mechanisms related with cloud formation that may be associated with the changes in the ionization of the stratosphere during the solar cosmic ray bursts [12]. With this in mind in this paper are presented several of the activities and recent results from BEO Moussala.

## 2. Atmospheric transparency measurements

The transparency is one of the primary measures of the atmospheric state. The precise long term series of atmospheric transparency measurements gives the possibility for quantitative estimate of the variability of air and therefore to make climatologic conclusions with regard to contamination, cloud formation, humidity and radiative exchange. Therefore it is very important to provide measurements of the integral atmospheric transparency.

It seems to be possible to estimate the atmospheric transparency on the basis of atmospheric Cherenkov light registration (Fig.2.), additional measurements with LIDAR and corresponding Monte Carlo simulations. The atmospheric Cherenkov light is produced by charged ultra relativistic particles in extensive air showers (EAS) [13]. The majority of the Cherenkov photons are produced near to the shower maximum [14]. Thus in practice the totality of the Cherenkov light passes trough the lower atmosphere. Obviously the atmospheric condition plays an important role and impacts the Cherenkov light propagation.

The registration of atmospheric Cherenkov light is possible using Cherenkov telescope. This device in our case represents system of two parabolic mirrors with focal length of 1.5m and diameter of 2m working in a coincidence regime (Fig.3.). Measuring the Cherenkov light flux produced in EAS in different atmospheric conditions one obtains different amplitude spectra. Obviously this reflects on the slope of the reconstructed spectrum. The different slopes of the reconstructed spectra correspond obviously to different atmospheric conditions. Thus it is possible to check the different absorption mechanisms measuring the atmospheric Cherenkov light. Moreover such type of experiment gives the excellent possibility to study the different atmospheric profiles using additional measurement with LIDAR or starlight extinction [19].

Generally two kinds of scattering are important: scattering by molecules of air, and scattering by solid particles or liquid droplets suspended in the air. Molecular scattering is usually called Rayleigh scattering. The suspended particles, on the other hand, are collectively known as aerosols, and their contribution is called aerosol scattering.

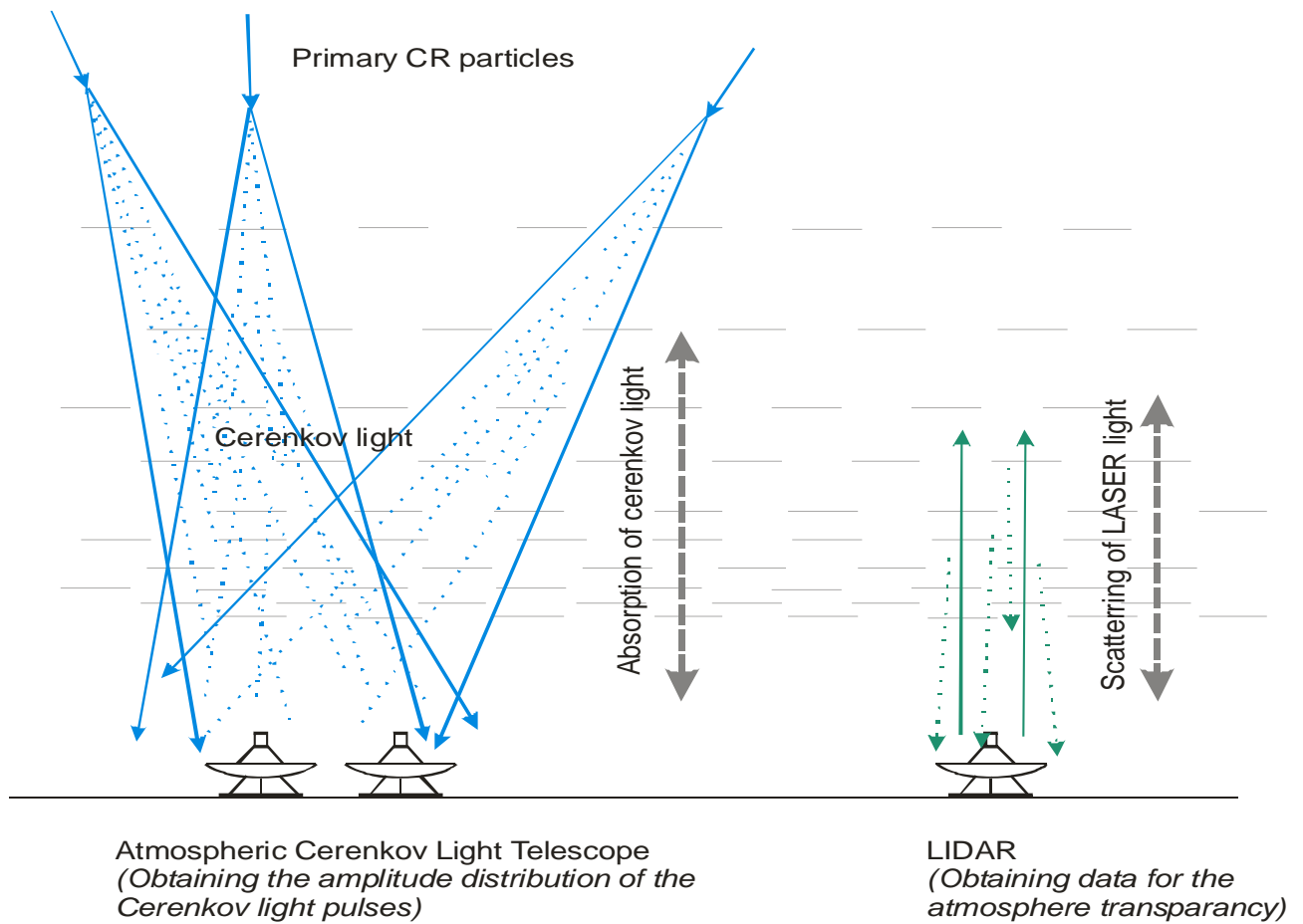

Fig. 2 Measuring atmospheric transparency on the basis of atmospheric Cherenkov light registration

In. Fig. 4 is shown the measured amplitude spectrum in cloudless condition. One can see the apparatus noise, the threshold and real counts used for reconstruction. It is important to compare such type of measurements with theoretical estimations or Monte Carlo simulations and of course with measurements based on different methods and techniques.

In this connection an additional Monte Carlo simulation using CORSIKA 6.002 code [15] with GHEISHA [16] and QGSJET [17] hadronic interaction subroutines have been carried out. The aim is in one hand to estimate the detector energy threshold and on the other hand to study the amplitude spectra in different atmospheric conditions. The simulated primary particles are protons with initial energy of $5 \times 10^{12}$ eV and distributed according spectrum with slope of 2.7. The American standard atmosphere was used for the simulations. The simulations were carried out in two cases – transparent atmosphere i.e. without any absorption or scattering of the light. The second case was with included Mie and Rayleigh scattering according [18].

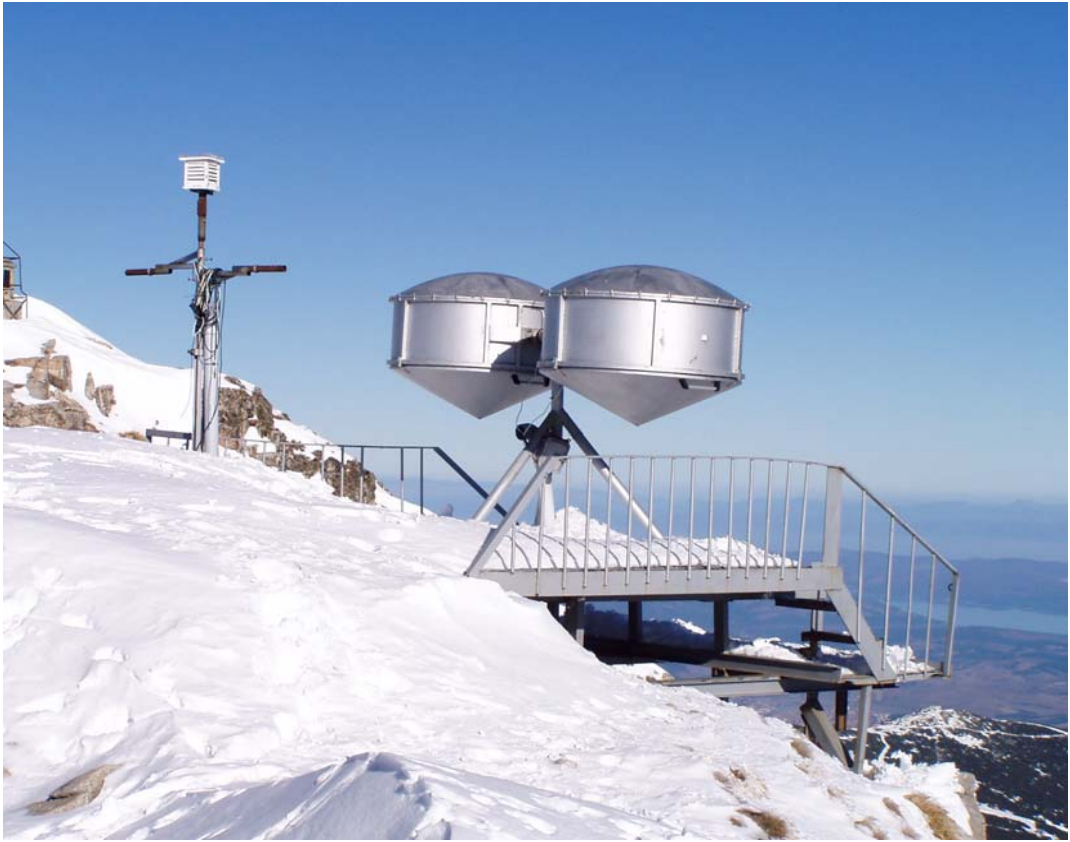

Fig. 3 Cherenkov light telescope at BEO Moussala

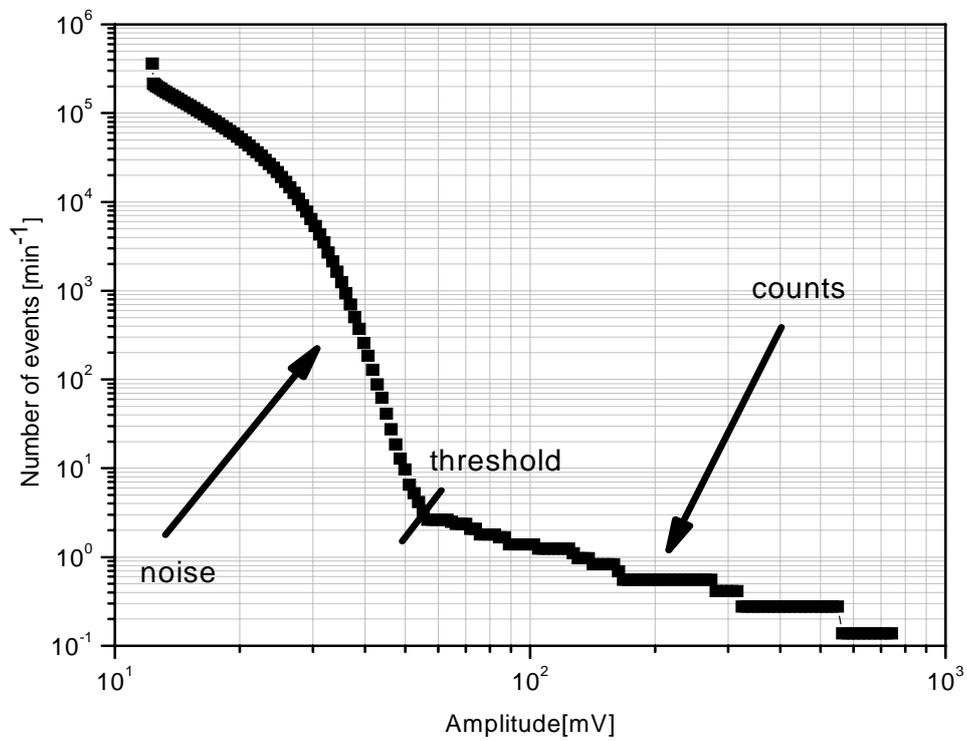

Fig. 4 Measured amplitude spectrum with Cherenkov telescope in cloudless conditions

The results of the simulation are presented in fig. 5. As a results the estimated energy threshold of the telescope is 5x10$^{13}$ eV energy of the primary particle which induce the EAS. The first experimental results confirmed the expectations concerning the different counting rates and slopes of the reconstructed spectra. However additional calculations will be needed. The aim is to obtain precisely the effective detector area as a function of the energy threshold of the telescope and different atmospheric conditions and profiles. This will permit to reconstruct the measured amplitude spectra precisely and therefore the obtained energy spectra slopes. Obviously additional measurements will be needed and comparison with LIDAR measurements in attempt to provide inter-calibration.

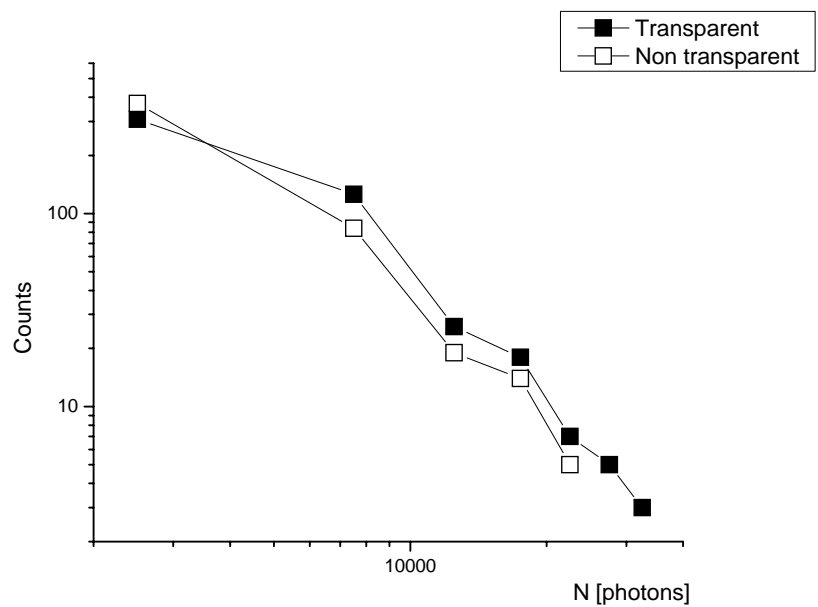

Fig.5 Simulated atmospheric Cherenkov spectra with CORSIKA code in the case of transparent atmosphere (black squares) and including Mie and Rayleigh scattering (open squares)

## 3. Neutron and muon flux measurements

As was mentioned the solar activity can affect the global climate in different aspects, timescales trough large diversity of mechanisms. Taking into account that the secondary cosmic ray neutrons and muons are connected with cosmic rat variations it is possible to study such effects on the basis of secondary cosmic ray muon and neutron flux measurements. In this connection aiming to study the variations of primary cosmic ray a neutron flux meter and muon telescope based on water Cherenkov detectors are developed at BEO Moussala. The aim of both devices is to investigate the possible connection between cosmic ray and climate changes [19].

The muon telescope is based on 8 water Cherenkov detectors. The water Cherenkov detector is a tank with 50x50x12cm (Fig.6). The principal aim is to investigate the variations of cosmic ray.

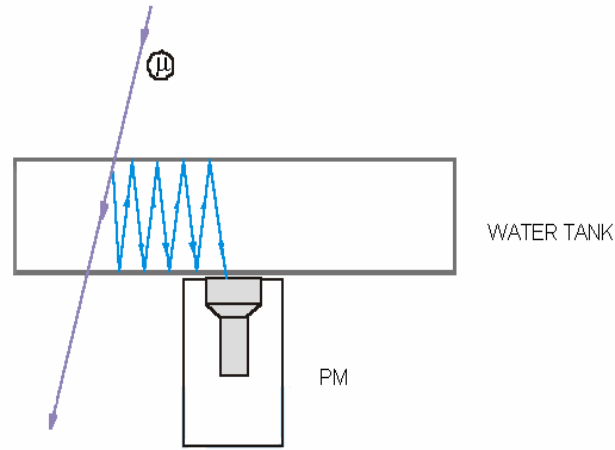

Fig.6 Water Cherenkov detector of the muon telescope

The tank efficiency registration was estimated with modified version of EGS4 code [20], precisely version with included Cherenkov effect [21]. Using the simulations with CORSIKA 6.002 code [15] with corresponding hadronic interaction models GHEISHA [16] for low energy interaction below 80 GeV/nucleon and QGSJET [17] for high energy hadronic interaction the distribution of muon component at BEO Moussala observation level was obtained. This permitted taking into account the concrete layer of the observatory to estimate the expected counting rate of the muon telescope for energy threshold of GeV. This permits to provide measurements with high statistics and study the cosmic ray variations. In Fig. 7 is shown the muon telescope at the basement of the observatory.

The recent results of the measurements are presented in Fig. 8 and Fig. 11. At the same time at BEO Moussala is operational automatic meteo-station Vaissala which gives information about atmospheric conditions such as pressure, temperature, wind direction and velocity etc…In Fig. 9 and Fig. 10 are presented the average of the atmospheric pressure and temperature for month of September 2006. Taking into account the results presented in Fig. 8 one can conclude that the muon telescope can provide precise measurements of the muon component of the secondary cosmic rays and thus gives the possibility to study the cosmic ray variation (detected pressure effect). Moreover registering the secondary cosmic ray muons it is possible to detect Forbush decrease. This permits to study the influence of galactic cosmic rays on the solar radiation input to the lower atmosphere, especially increases of the total radiation fluxes associated with Forbush-decreases in the galactic cosmic rays [22].

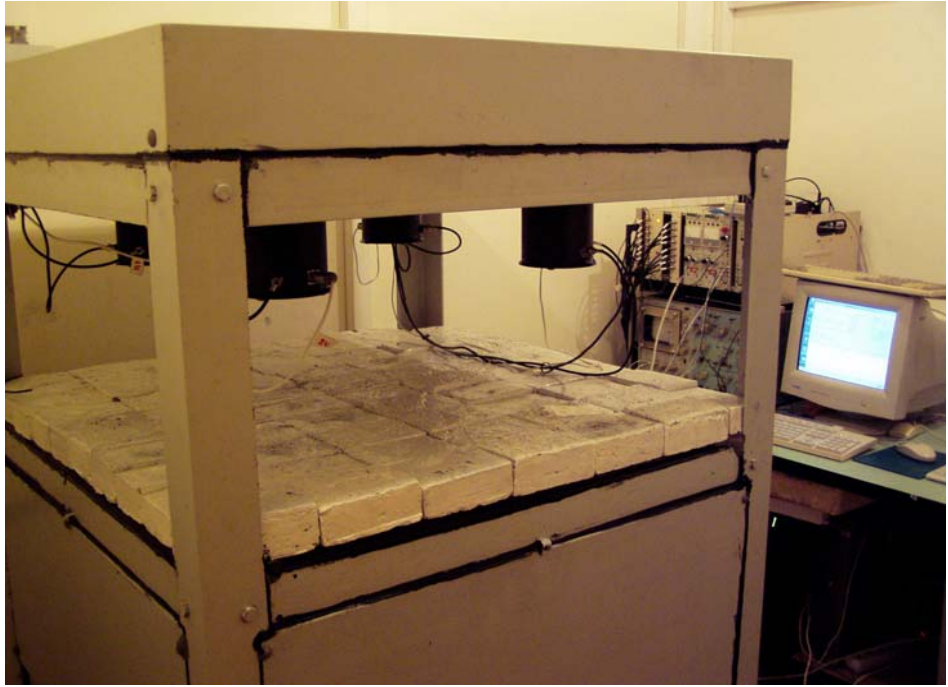

Fig. 7 Muon telescope based on water Cherenkov detectors

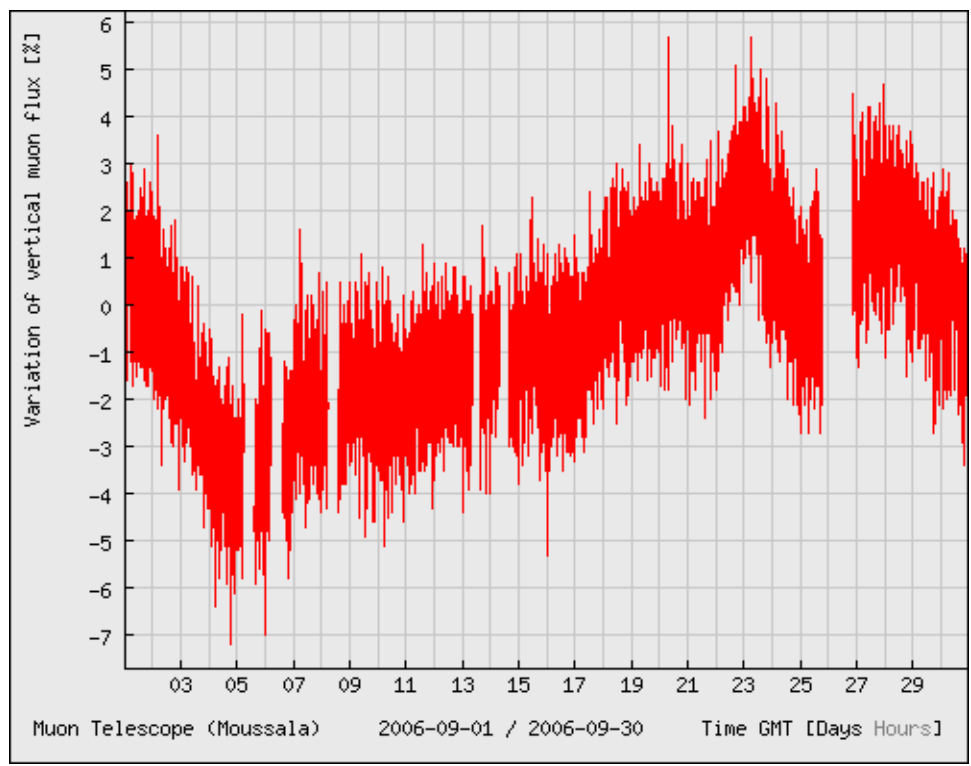

Fig. 8 Variation of vertical muon flux measured by muon telescope for September 2006

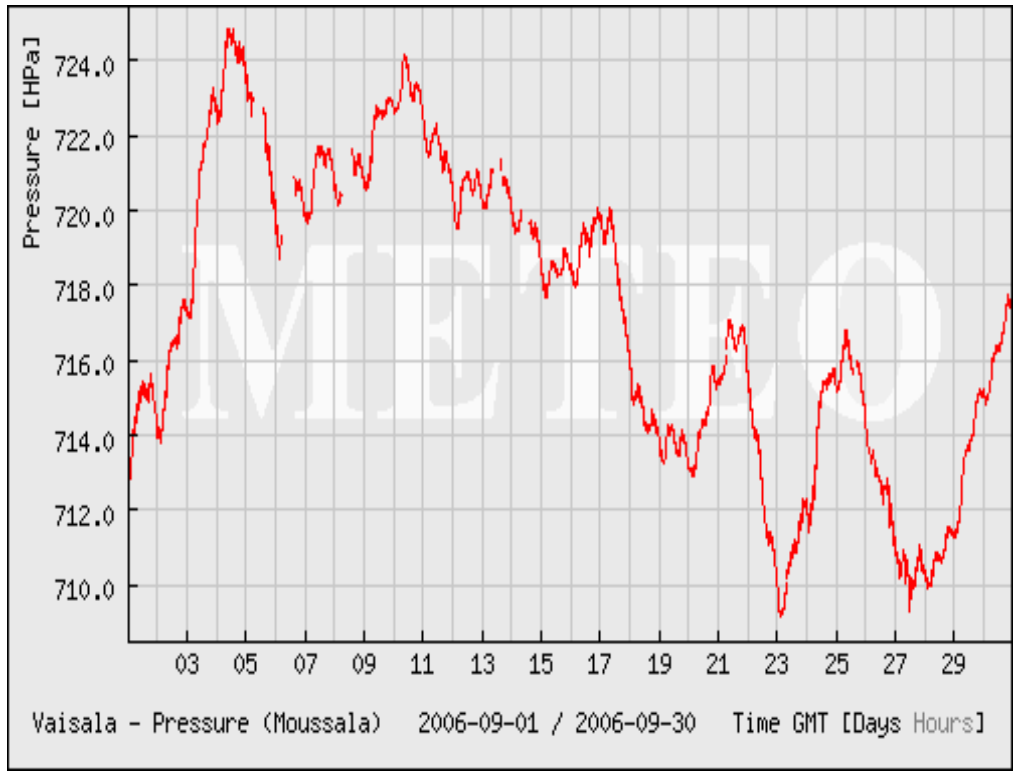

Fig. 9 Average of the atmospheric pressure at BEO Moussala for September 2006

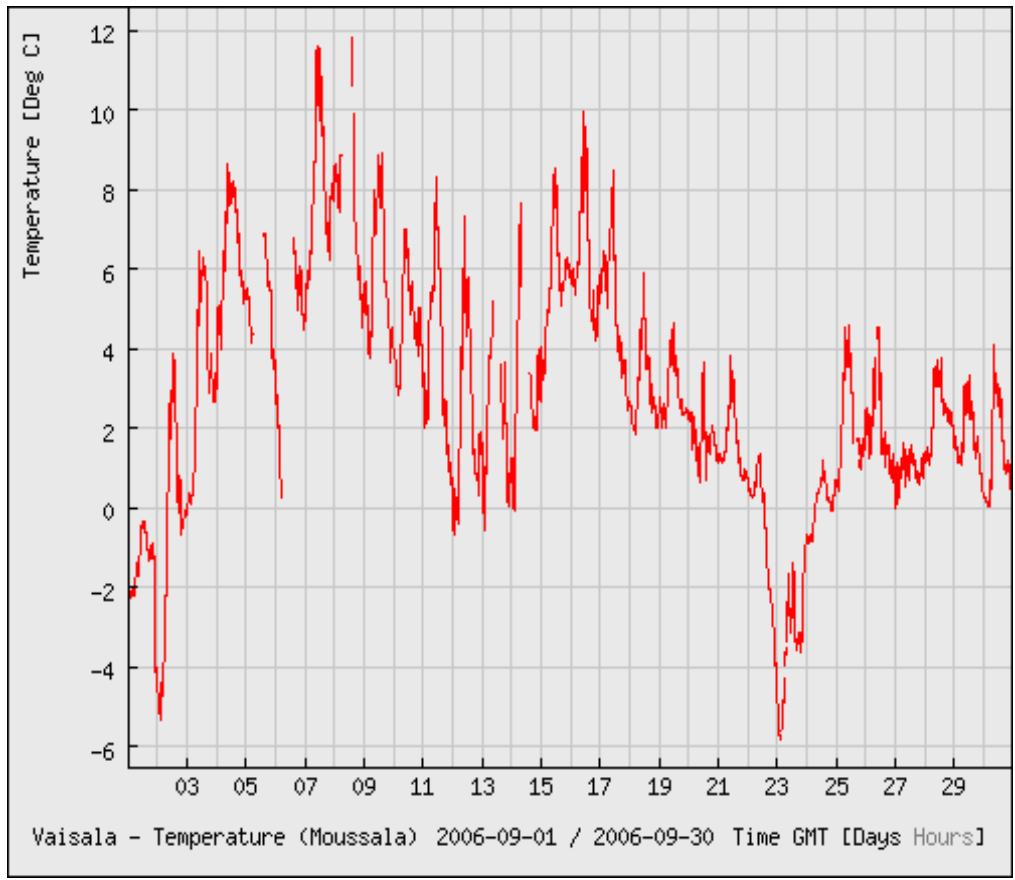

Fig. 10 Average of the temperature at BEO Moussala for September 2006

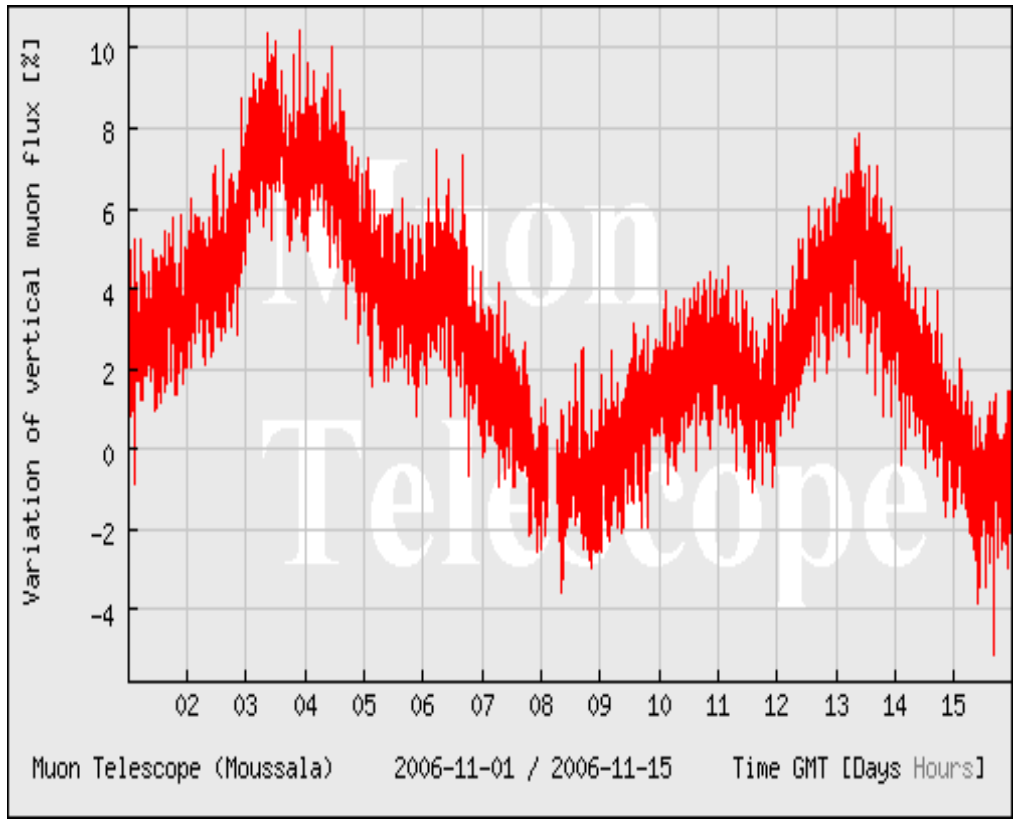

Fig. 11 Variation of vertical muon flux measured by muon telescope for November 2006

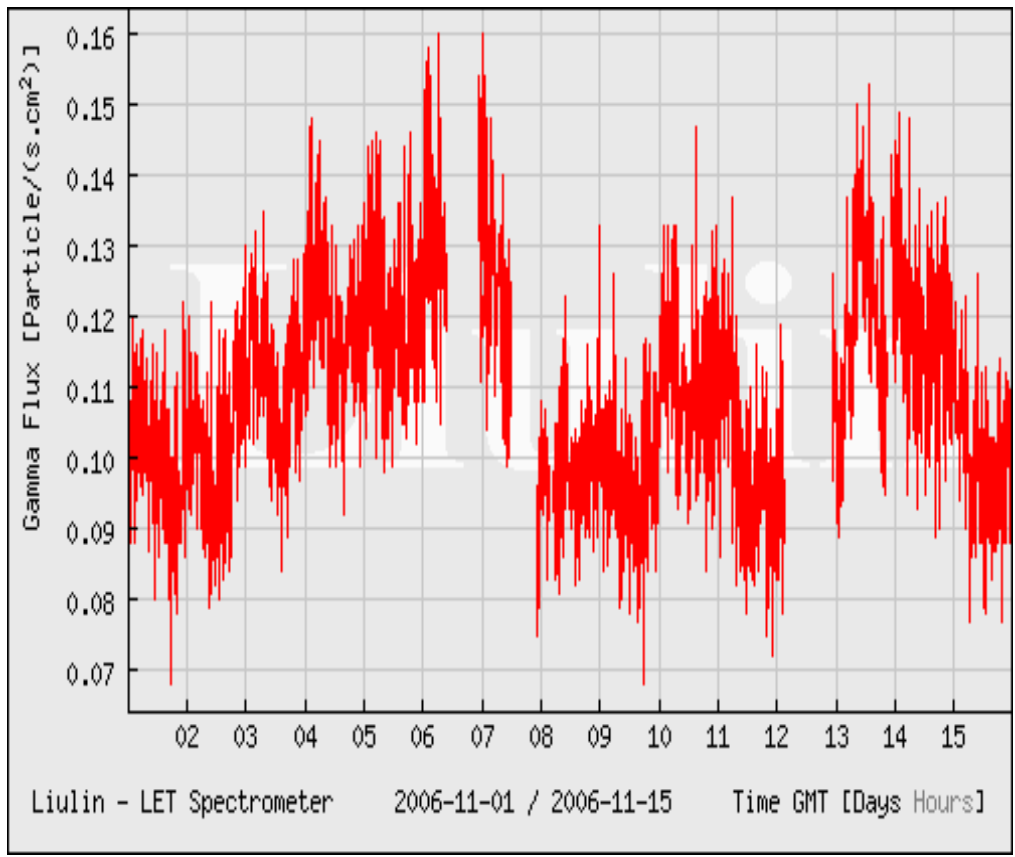

Fig. 12 Variation of particle flux measured by Liulin LED spectrometer for November 2006

Additional measurements with Liulin LED spectrometer of the secondary cosmic ray particles flux shows good correlation with the muon telescope data (Fig. 12).

Complementary to muon telescope a neutron flux meter is developed. It represents complex of six gas filled detectors type SNM-15 with $BF_3$ enriched to 90% with $B^{10}$. The detectors are situated under the roof of the main building of BEO Moussala. The detector complex is divided in two modules of 3 detectors. In Fig. 13 is presented the front panel of the first module including the signal and power supply cables. The device is without lead i.e. is only with neutron moderator - glycerin. This is the main difference comparing to the usual neutron monitors. According the initial design [23, 24] the aim of the complex is to provide with high statistics and precision measure of the absolute secondary neutron cosmic ray flux. The scientific potential of detector complex is enormous. The relativistic cosmic rays both galactic and solar play a useful key in space weather storms forecasting and in the specification of magnetic properties of coronal mass ejections, shocks and ground level enhancements [25].

Moreover it is possible to investigate the variations of the pressure level heights, temperature profiles and wind characteristics in the troposphere and lower stratosphere during Forbush-decreases of the galactic cosmic rays [26]. The Forbush-decreases are accompanied by the pressure increase in the whole troposphere, the maximum of the effect taking place on the 3–4th day after the event onset. Simultaneously the temperature decrease is observed in the troposphere during the first few days of the Forbush-decreases. The pressure increase might be related to the changes of wind characteristics in the middle and upper troposphere. A possible mechanism of the observed effects seems to involve radiation budget changes in the atmosphere due to the cloudiness variations associated with Forbush-decreases of the galactic cosmic rays.

At the same time the precise measurements with both mentioned above devices gives excellent possibility to understand the role of cosmic ray variation of the Earth climate and to check different mechanisms of such type of influence [27, 28].

The detailed Monte Carlo simulation of the detector response was carried out. The result is the estimation in one hand the moderator layer of glycerin (12.5 cm) and on the other hand the expected counting rates at Moussala observation level of 2925m above sea level (725 g/cm2). In this study was used the measured at Testa Griggia neutron spectrum [29] (Fig. 14). In fig. 15 are presented the results of simulation with MCNP(x) as a neutron total current function of the moderator layer. The 12.5 cm moderator of glycerin permits in one hand to avoid the registration of neutrons result of reaction near to detector and on the other hand to assure relatively high counting rate of the detector complex and thus to provide good statistics of the measurements.

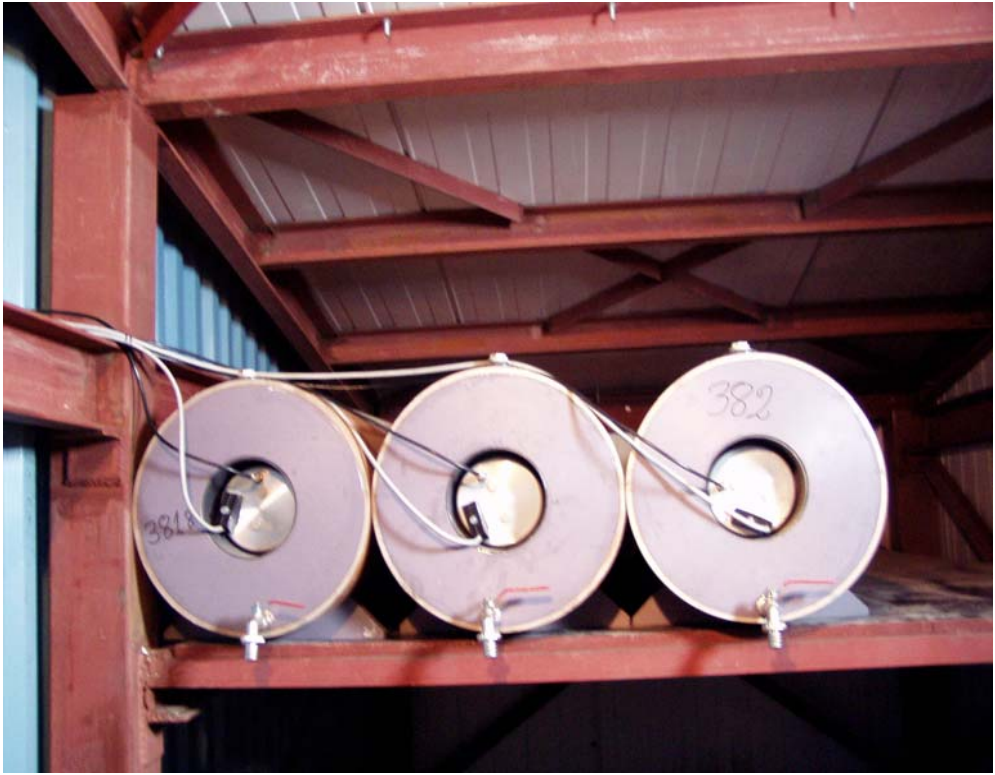

Fig. 13 Front panel of the neutron flux meter at BEO Moussala

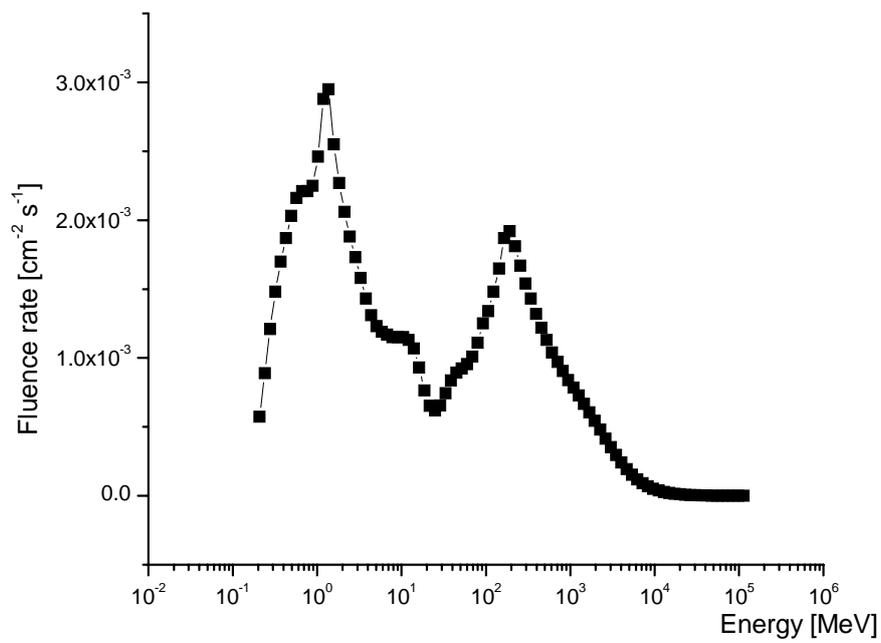

Fig. 14 Testa Griggia neutron spectrum [29]

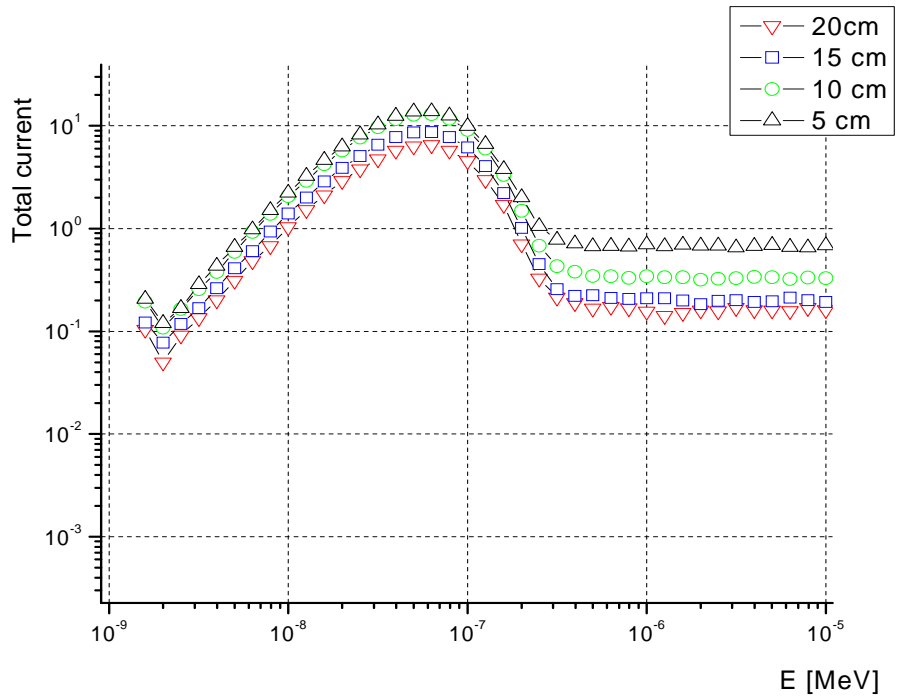

Fig. 15 Simulated total neutron current

The preliminary measurements at Sofia observation level – 550 m above sea level, which corresponds to 970 g/cm$^2$ and Moussala observation level taking into account the detector efficiency [30] and more realistic geometry shows very good coincidence of the measured and simulated counting rates. The estimated counting rate for Sofia observation level is around 3 neutrons/s for single detector as a total neutron current. The complex of 6 detectors gives an average of registered events approximatively 12 for 10 minutes measurements. Taking into account the estimated registration efficiency and secondary cosmic ray neutron spectrum at Sofia observation level one obtains very similar counting rate for simulated and measured rates.

The geomagnetic and radiation storms are significant elements of space weather [31]. The forecasting of such type of events is very important for orbiting flights. In fact the geomagnetic storms are driven by magnetized plasma clouds. They reach the Earth from few hours till several days. During their propagation they interact with galactic cosmic rays. The result is the modulation of galactic comic rays till energies of thousands of GeV. As was mentioned above the change of the intensity is possible to detect by surface monitors. Additionally muon detector network [32] can provide powerful tool for better understanding the space weather in the vicinity of Earth. At the same time sudden correlated measurements and analysis of the variation of secondary muon, neutron and electrons could be good basis for indication of upcoming geomagnetic storms.

Therefore for real time studies of solar-terrestrial relations it is required to register simultaneously as more as possible of phenomena in heliosphere. In this connection it may be very useful to use ground-based muon hodoscope with high angular resolution that detects muons of cosmic rays with energy around 10GeV. Solar flares, scattering of protons by interplanetary shock waves, fluctuations of the air density distribution in the atmosphere will change ground level muon intensity. Amplitude of such variations can reach maximal value in various energy-active regions of the Earth near magnetic poles, tropics, sea coast of continents etc. In this connection a good solution for BEO Moussala will be the development of Nor-Ambered multidirectional muon monitor [33] see Fig. 16. The detector complex consists of two layers of plastic scintillators below and above of the sections of the Nor Ambered neutron monitor. The lead absorbs the low energy muons and the electronic component. The estimated threshold is around 350 MeV for detected muons.

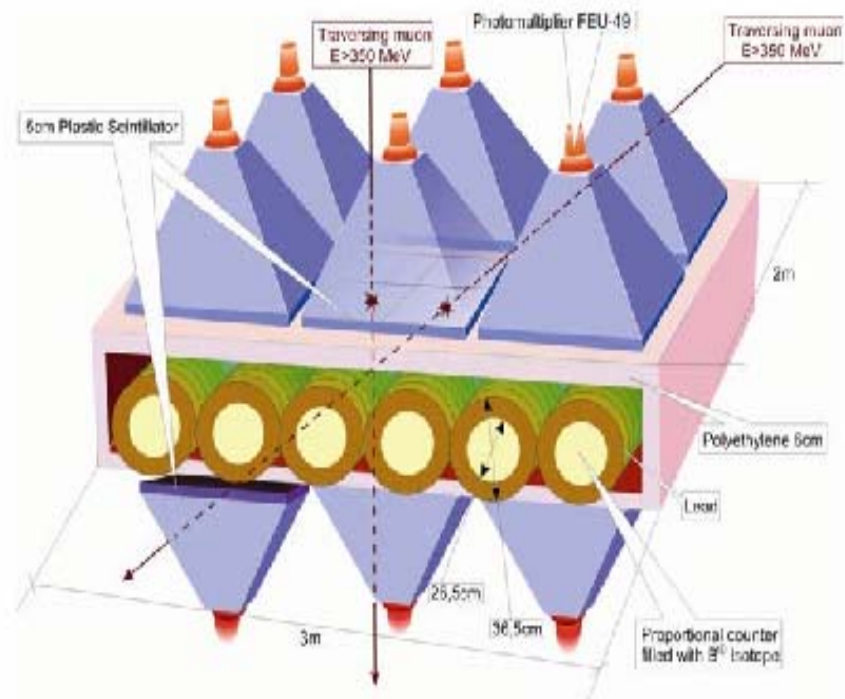

Fig. 16 Nor Amberd multidirectional neutron monitor [32]

An additional device which will be complementary is the muon hodoscope (Fig. 17). According the initial design [34] the muon hodoscope represents multi-channel device. The complex is based on water Cherenkov detectors [19, 30] with cylindrical form and dimensions 10 cm diameter and 1m length. The muon hodoscope is made of four layers with 128 counters in a layer. The distance between the two pairs of layers is 1 meter. The estimated registration efficiency is 91% assuming 5 GeV energy muons for threshold and using model [21]. The scientific potential is enormous starting from internal gravitational waves, measurement of the temperature field along height of the atmosphere, registration of acoustic waves [19] etc…

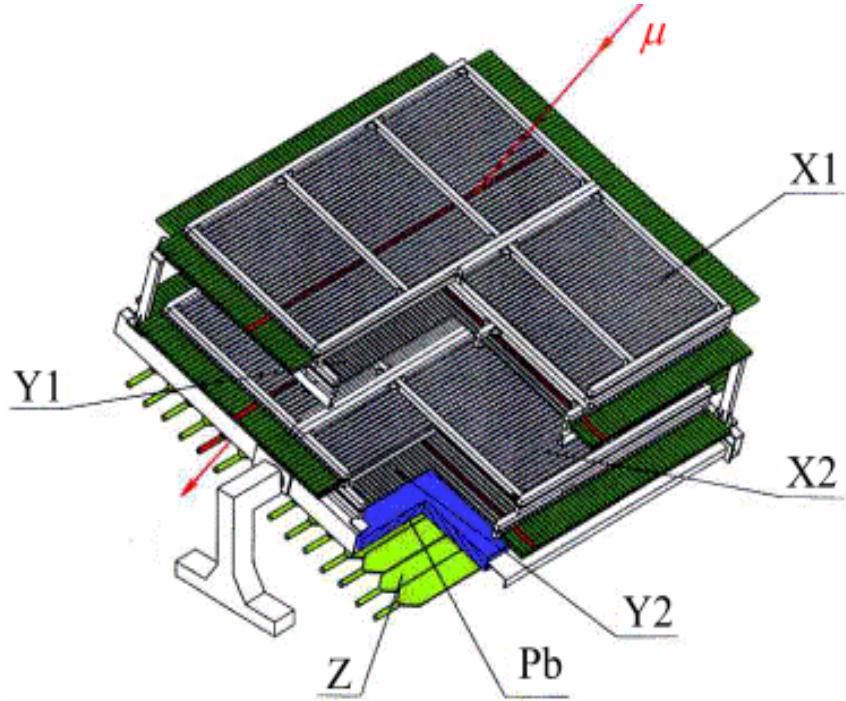

Fig. 17 General view of the muon hodoscope

## 4. Summary

In this work are presented several at BEO Moussala activities connected with secondary cosmic ray registration, especially the atmospheric Cherenkov light, neutron and muon component. The presented activities are both theoretical, precisely estimation of the energy threshold, efficiency registration based on Monte Carlo simulations and experimental. Several preliminary results and estimations based on Monte Carlo simulation with CORSIKA code were carried out in attempt to obtain the energy threshold of the atmospheric Cherenkov telescope. As example is shown measured amplitude spectrum. The future plans are connected especially with more precise and detailed simulations of different components of EAS at Moussala observation level of 725 g/cm$^2$. This will permit in one hand to obtain the effective area of the atmospheric Cherenkov light telescope and therefore to provide more precise analysis of the measured events. On the other hand this will permit to make detailed

Monte Carlo simulation of the detector response of particle detectors. All these results are good basis for building reconstruction strategy.

The muon telescope based on water Cherenkov detectors is shown. The registration efficiency of single tank is obtained and the whole device efficiency is estimated. Several experimental results are shown.

The neutron flux-meter is described with estimations concerning the final design. The Monte Carlo simulations and measured counting rates for Sofia observation level are compared. The Nor-Ambered multidirectional muon monitor and muon hodoscope project proposal was mentioned. The scientific potential of all devices is discussed. It is clear that the high quality data can be useful as a basis to check several models on the influence of high energy particles on the middle atmosphere. A good example is presented in [19].

**Acknowledgements**

We acknowledge the BEO Moussala staff and our colleagues from Lomnicki Stit observatory especially Prof. K. Kudela. We warmly acknowledge prof. M. Storini from IFSI/CNR Italy for the fruitful discussions during BEOBAL workshop in September 2005. We warmly acknowledge prof. E. Eroshenko from IZMIRAN and colleagues from Solar Terrestrial influences laboratory prof. P. Velinov and Y. Tassev for the suggestions concerning the neutron flux measurements and future plans at BEO Moussala. This work is supported under NATO grant EAP. RIG. 9811843 and FP6 project BEOBAL.